# Process Model Difference Analysis for Supporting Process Evolution

Martín Soto, Jürgen Münch

Fraunhofer Institute for Experimental Software Engineering,
Fraunhofer-Platz 1, 67663 Kaiserslautern, Germany
{soto, muench}@iese.fraunhofer.de

**Abstract.** Software development processes are subject to variations in time and space, variations that can originate from learning effects, differences in application domains, or a number of other causes. Identifying and analyzing such differences is crucial for a variety of process activities, like defining and evolving process standards, or analyzing the compliance of process models to existing standards, among others. In this paper, we show why appropriately identifying, describing, and visualizing differences between process models in order to support such activities is a highly challenging task. We present scenarios that motivate the need for process model difference analysis, and describe the conceptual and technical challenges arising from them. In addition, we sketch an initial tool-based approach implementing difference analysis, and contrast it with similar existing approaches. The results from this paper constitute the requirements for our ongoing development effort, whose objectives we also describe briefly.

## 1 Introduction

Software development organizations striving to achieve a high level of process maturity must sooner or later face the problem of process standardization, namely, guaranteeing that all organization units develop software according to one well-known, unified process. Achieving process uniformity generally requires the definition of *standard processes* (sometimes also called reference processes or generic processes) that capture organization-wide process knowledge, possibly with emphasis on a particular application domain (e.g., space software) and/or on specific development contexts (e.g., large projects). However, since they are generic, standard processes must be tailored to the particular needs of the various projects inside the organization, leading to many separate *project-specific processes*.

Both standard and project-specific processes are subject to evolving along their life cycle. Rapid technology changes, newly available useful knowledge, changes in regulations or process standards, and new project experience, to only mention a few factors, contribute to push processes in different directions. Moreover, processes need to be designed, described, introduced, and maintained in such a way that they become accepted by practitioners and thus actually used in practice. For this reason, evolution must be guided by solid, practical experience.

The problem of driving process evolution based on experience involves activities both at the organizational and at the project level. Initially, particular projects tailor processes to their needs and proceed to enact them. During enactment, issues involv-ing the process definition are typically observed, ranging from the need to refine certain process entities in order to make them more specific, to the identification of areas of the process definition that are openly inadequate and must be redefined.

Incorporating this local, project-specific experience into the standard organizational process is a potentially complex task involving at least the following two steps. First of all, local variations must be identified and characterized in order to determine if they are general enough to become part of the standard process. Afterwards, selected local variations must be generalized and added to the standard process as alternatives, together with constraints or rules limiting their use to particular cases. This, of course, requires a deeper understanding of the appropriateness of the process alternatives for different contexts and their effects on these contexts.

Additionally, before the start of a new project, a characterization of the project context and its goals must be produced, providing the information needed to select adequate process alternatives for the project. This closes the experience cycle, opening opportunities for experience reuse.

We believe that the first step can be effectively supported by so-called *process model difference analysis*, namely, finding, analyzing, and displaying the differences between variants of a single process model in ways that are meaningful, and thus useful, to the people maintaining and using the process. The second step addresses the so-called *variability analysis*, i.e., identifying which context characteristics and project goals differ among a family of projects, and determining the corresponding process variation points and the rules associated to them. The concept of variability analysis originally comes from product line engineering [1].

This paper presents our current steps towards an effective, practical approach for process model difference analysis. The rest of the paper is structured as follows: In Section 2, we present two process management scenarios derived from our experience with process modeling and implementation, analyze the possible role of difference analysis in them, and derive a set of basic interesting difference analysis operations. In Section 3, we discuss the conceptual and technical challenges of process model difference analysis, and contrast them to existing procedures like the standard longest common subsequence algorithm used by diff. Section 4 discusses the basic concepts of our ongoing implementation work. Section 5 presents some related work and Section 6 concludes the paper by discussing open challenges and plans for realizing our view.

## 2  Application Scenarios for Difference Analysis

In the following, we sketch two scenarios that demonstrate the need for process model difference analysis. These scenarios are based on the authors' experience in defining and managing the evolution of process standards (such as the SETG [2] of the European Space Agency) and implementing compliance management in organizations. The scenarios are used to identify a set of basic operations involved in difference analysis. For each one of the two scenarios, we describe the problem at hand and identify the process stakeholders (or rather, stakeholder roles) involved in it. In a second step, we list the questions that each stakeholder must answer in the context of the scenario, together with the difference analysis operations that can be used to support the stakeholders in answering these questions.

## 2.1 Scenario 1: Definition and Evolution of Process Standards

In principle, there are two main approaches to the definition of process standards: *top-down* and *bottom-up*. In the top-down approach, a standardization board collects individual experiences, methods found in literature, or requirements enforced by other standards, and creates a prescriptive process model, which is then provided to the development organization and empirically optimized later on. The ECSS [3] standards for space software, or the German national V-Modell XT standard [4] are examples of the top-down approach. In the bottom-up approach, standards are mainly developed based on observation and descriptive modeling. The WISEP reference process for wireless Internet services [5] and the LIPE reference model for e-business software development [6] illustrate this approach. It is important to observe that, independently of how process evolution is managed, observing processes in practice, identifying variations in them, analyzing these variations, and feeding them back into the standard process model [7] are fundamental activities for actual improvement. This feedback cycle can be supported by process model difference analysis.

One typical scenario is that a large software organization distributes a single process model to several of its development units, which is intended to be used as the main software process description for conducting independent software development projects. Since the defined process has not been widely tested in the context of the organization, and since conditions differ from one project to the next, individual projects are allowed to adapt the process description in an ad-hoc manner to better suit their particular needs.

After a few months, the independently tailored process models have diverged significantly. This poses a number of challenges:

– The central organization wants to make sure that, despite project differences, a unified basic process is followed by all projects, and that the customization of this process is done in a systematic way. In other words, it is important to prevent local processes from diverging too much from the established organization standard.
– Additionally, practices introduced by individual projects may turn out to be useful to other projects. It would be valuable to identify such practices, abstract them, and eventually integrate them with the generic organization-level process definition.
– Furthermore, it would be valuable to identify areas of the current process that adequately fit the organization's environment, as well as areas that may be difficult to enact in the current environment. It would also be important to identify areas that, although adequate, may require improvements in their documentation.
– Software managers, software developers and, generally, personnel working on software projects, may be moved between projects based on changing organizational needs and priorities. People used to one project's process definition may have problems getting acquainted with new, slightly different processes between their previous and new projects. Process difference analysis could help to identify these differences and provide guidance for working in the new project.

A similar scenario arises when a reference process model (e.g., V-Modell XT or ECSS) is adopted and further tailored by separate organizations. The standards body responsible for the reference model may be interested in collecting feedback from process users in order to determine how the reference model should evolve.

The following table lists involved stakeholders, their questions, and the way process model difference analysis can support them in answering their questions:

| Stakeholder | Question | Helpful difference-analysis operations |
|---|---|---|
| Software Process Group | Are there any structural changes (new/deleted activities/products, different relations) in project processes with respect to the organization's process? | Visualize structure with differences. |
| | Do structural changes affect the general process structure or only the detailed structure of particular process areas? | Provide different views into process structure and structural differences: general, per process area, per role, etc. |
| | Which entity descriptions were modified? What sort of modifications happened? | List changed descriptions. Highlight entities in the general structure whose descriptions changed. Measure the extent of changes and visualize it based on the structure (i.e., map trees.) Apply text comparison to descriptions. |
| | Which areas of the process were changed by many projects? Are the changes similar? | Present differences with respect to the main model in parallel. Apply similarity detection algorithms to common changed areas. |
| Project Manager | Which process changes have we made until now? Can we justify them based on our concrete project needs and requirements? | Visualize structural differences, including views. Visualize description differences on top of the structure. Visualize recorded rationales for changes [17]. |
| Developer (process agent) | What is different between the process I used to follow in my old project and the process defined for my new project? | Compare processes from the old and the new project with common ancestor (main organizational process is the ancestor.) |
| | What's special in my new project's process with respect to the general organization's process I learned in my training? | Compare process with ancestor. |

### 2.2 Scenario 2: Process Compliance Analysis

Nowadays, more and more organizations are subject to regulatory constraints requiring the existence of explicit processes, as well as adherence to them (see, for example, the IEC 61508 standard for safety-related systems [8].) Being compliant typically requires maintaining traceability information that captures the relationships between the actual and the prescribed development processes, a difficult task since, for a variety of reasons, it is possible for both models to evolve, thus leading to deviations. Difference analysis can help to characterize the evolution in order to determine whether action is necessary to stay compliant. In addition, traceability information needs only to be updated for those process parts of the models that changed.

The following is one typical scenario: A development organization adopts a reference model as a base definition for its development processes. As usual with reference models, although they provide a good framework for process definition, some aspects of them must be adapted to the unique needs of each organization. For this reason, a tailoring effort is launched, which concludes several months later with a process definition adequate for being used by new development projects at the organization. Some time afterwards, and independently from all internal process efforts, a new version of the reference model is published. There is pressure from inside and outside the organization to use this new version of the reference model. However, the organization does not want to lose the significant effort invested in tailoring the old version. The transition poses a number of difficulties:

- It is hard to determine which tailoring changes can be moved to the new version of the reference model directly, which of them can be adapted, and which must be discarded because either they are now covered by the new model or they conflict with it.
- Moreover, since it is difficult to reliably identify the areas that must be changed, even estimating the effort necessary to produce a tailored variant of the new reference model version can be very hard.
- In addition, standardization organizations typically do not give sufficient information about the detailed changes. Often, differences between new versions are only described on an abstract level (e.g., the new standard focuses more on reliability), but it is unclear which process elements have changed.

The following table is similar to the one included in the previous scenario:

| Stakeholder | Question | Helpful difference-analysis operations |
|---|---|---|
| Software Process Group | How exactly were the structure and contents of the reference model modified? Which actual elements were affected and how? | Compare process with ancestor (old version of the reference model is the ancestor.) Visualize structure with differences. |
| | How exactly did we tailor the structure and contents of our current process model? Which actual elements were affected and how? | Compare process with ancestor (old version of the reference model is the ancestor.) Visualize structure with differences. |

| Stakeholder | Question | Helpful difference-analysis operations |
|---|---|---|
| | Which areas did we tailor that remained essentially untouched in the new reference model version? Which areas were modified in the reference model that we did not touch? Which areas were changed in both cases (conflicts)? | Compare processes with common ancestor (old version of the reference model is the ancestor.) Visualize structure with differences. |
| | How big are the conflicts? Were do the most complex conflicts lie? | Measure the extent of changes. Compare and visualize. |
| | Are there structural or content related similarities between our changes and the changes made to the reference model? | Apply similarity algorithms to selected portions of the model. |

### 2.3 Further Applications

Analyzing and visualizing differences between process models can be used in many other situations: An example application is the collaborative design of development processes. Here, difference analysis can be used during the integration of parallel designed processes. Another example for the use of process difference analysis is the development of systems for process versioning and configuration management. Here, differences between process models can be determined and used as deltas to calculate previous versions of process models.

## 3 Difference Analysis Challenges

Based on the set of useful operations presented above, this section discusses the main challenges we observe in process difference analysis. These challenges cover various conceptual and implementation issues.

### 3.1 Filtering and Presenting Results for a Multitude of User Groups

Practical process models used in real-world development organizations are often very complex, comprising a large number of interrelated process entities (activities, artifacts, roles, etc). For this reason, a large majority of process stakeholders have to deal with only one portion or aspect of the process model (e.g., only the analysis or the testing process; only administrative or technical portions of the process; only high-level process descriptions; etc.) while performing their daily work.

As shown in the scenarios, the need arises to provide such users with difference analysis operations that are particularly tailored to their needs. This requires a flexible notation for specifying comparisons that is able to express the composition of a variety of filtering, transformation, and visualization algorithms, among other possibilities, to produce the difference analysis results.

Figure 1 shows a graphical comparison of two variants of a hierarchical structure (for example, an activity hierarchy in a process model) that we kept intentionally small for illustration purposes. Such a difference analysis would require filtering the model variants to extract the desired hierarchy, comparing them, and producing an adequate visualization with a graph layout algorithm.

### 3.2 Genericity

Our experience shows that organizations tend to have very specific, idiosyncratic ways to speak about software development and software development processes. Even if the general concepts used to model software processes tend to be similar, the way they are exactly defined as well as the terminology used to refer to them may vary widely among different software organizations, or even between divisions of a single organization.

Such a variety of process model schemata further complicates difference analysis. Even if we do not try to support comparing models structured according to different schemata, comparison must often make use of schema information in order to produce meaningful results. For example, particular attributes (e.g, long text descriptions) of certain entities belong to data types that require comparison with specialized algorithms (e.g., LCS-based text comparison). Also, the model may contain portions that, based on the schema, may be known to correspond to sequences, trees, or some other known structures that can benefit from being processed with more specialized algorithms.

### 3.3 Multiple Comparison Algorithms (or, Why diff is Not Enough?)

Comparing source code versions and analyzing the resulting differences (often referred to as *patches*) is a task software developers perform on an almost daily basis. Source code comparison serves a variety of purposes, like sharing of changes; review and analysis of changes done by others; space-savvy storage of multiple versions; and measurement of the extent and scope of changes; among others. Such comparisons can be performed using widely available software, like the well-known Unix diff utility, and similar programs.

An obvious question when speaking about model difference analysis is whether the problem is not solved by just storing the models in files and comparing them using diff. Although this is usually possible, it is almost always the case that the results delivered by diff are practically unusable. Diff relies on interpreting files as being composed of text lines (sequences of characters separated by the newline character) and then finding the *longest common sequence* (LCS) of lines by using an efficient algorithm (see [9] for example). The underlying practical assumption is that the material in the file can be read and understood sequentially.

Although this assumption holds true for source code files, process models usually follow patterns that resemble trees or, more generally, graphs instead of plain sequences. They are often heterogeneous in nature, being composed of pieces of data that follow different structural patterns and are represented in diverse ways. Of course, it is always possible to use LCS-based algorithms to compare certain portions of a process model (like text descriptions). It is also possible to store complete models in a line-oriented format (i.e., a text-based formal process model notation) and compare that representation. Although such an approach can be useful for determining differences in particular denotations of a model, we deem it insufficient to cover the wider range of abstract, task-oriented comparisons we are considering.

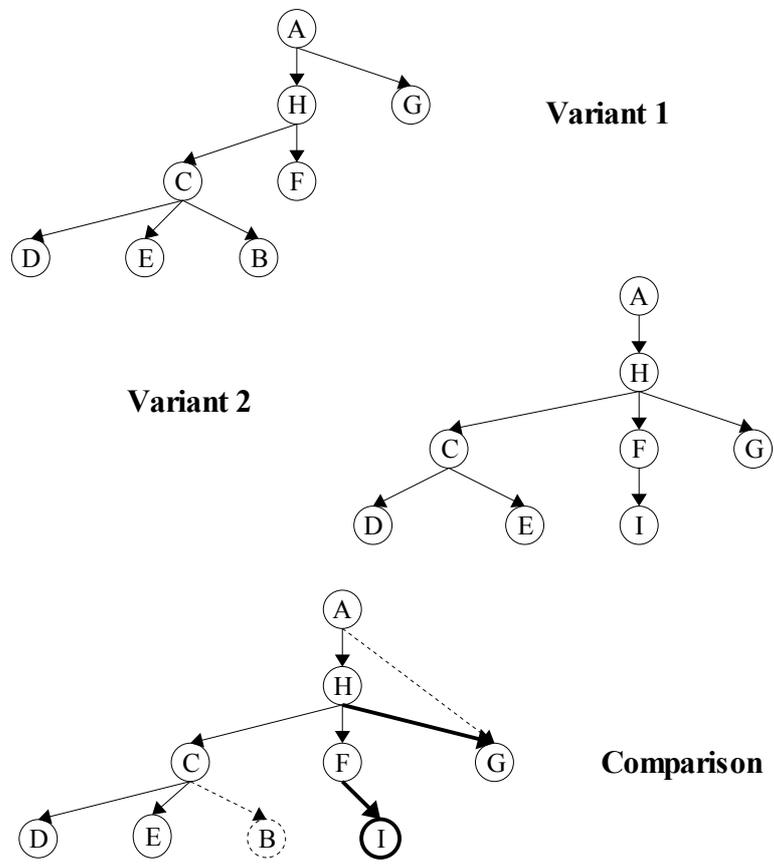

**Fig. 1.** Hierarchy difference analysis. The first two graphs represent variants of the same hierarchy (for example, with nodes corresponding to process activities and arrows corresponding to a *has-subactivity* relationship.) The third graph displays the differences between the two hierarchies: dashed elements are present only in the first variant, whereas elements drawn in bold only appear in the second one. Other elements are common to both variants. Such a display can be very useful for quickly identifying differences between complex structures.

### 3.4 Detailed Change Histories versus Difference Analysis

It is also possible to determine version differences along the evolution of a process model by simply recording every change as it is done. Keeping such a change log manually, however, is very hard, unreliable work that often prevents people from concentrating on their main tasks. For this reason, the only viable alternative is to embed support for recording changes in process modeling tools (similar to the "track changes" function available in common word processing programs).

Even if that is the case and although such change traces can be useful for certain purposes (e.g., auditing) they often contain too much information for most other purposes. For example, changes must often be undone, or they get superseded by larger modifications. Most difference analysis users are not interested in such minutiae. Proper difference analysis requires expressing the differences in a condensed, targeted form, which frequently can be obtained by directly processing the models instead of looking at their detailed change history.

## 4 A Preliminary Architecture for Difference Analysis

At the time of this writing, we are taking the first steps to produce a practical implementation of the vision presented in the previous chapters. In this section, we briefly discuss the elements that, according to our current vision, should comprise an adequate process model difference analysis system.

A block diagram for our architecture is shown in Figure 2. It is comprised of the following components:

– A *model importer,* which purpose is reading model variants in diverse formats and storing them in a common, comparable format in the model database.

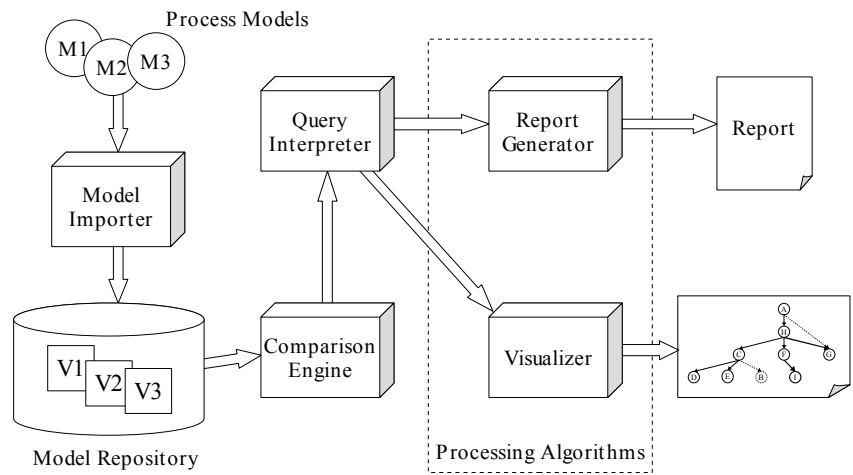

**Fig. 2.** Block diagram for a preliminary difference analysis architecture.

- A *model database,* containing a number of model variants. The database stores process models using W3C's Resource Description Framework (RDF) [18] as a generic notation. RDF is able to represent internal model structures like graphs, trees and sequences. Data attached to such structures, like text descriptions and graphics, can also be stored as RDF literals. Currently, we are testing a trial implementation of such a database, based on a standard relational database system.
- A low-level *comparison engine*, which calculates raw differences between model variants. This engine takes two variants of a model and produces a single model (called the *comparison model*) that contains the elements from both variants decorated to indicate whether they are common to both variants or exclusive to one of them. Our intent is to also use RDF to express such unified comparison models.
- A specialized *query language interpreter*, able to direct the above engine to build a comparison model from two given model variants, and further filter and process it in a variety of ways. This language is also able to feed the (potentially filtered) comparison model to other algorithms for further processing or visualization.
- A number of *visualization and display algorithms* intended to provide a high–level view of the comparison results.

## 5 Related Work

Although no previous work we know about specifically deals with analyzing and visualizing differences between process models, other research efforts are concerned in one way or another with comparing model variants and providing an adequate representation for the resulting differences.

[10] and [11] deal with the comparison of UML models representing diverse aspects of software systems. These works are generally oriented towards supporting software development in the context of the Model Driven Architecture. Although their basic comparison algorithms are applicable to our work, they are not concerned with providing analysis or visualization for specific users.

[12] presents an extensive survey of approaches for software merging, many of which involve comparison of program versions. Most program comparison, however, occurs at a rather syntactic level, and cannot be easily generalized to work with more abstract structures like process model graphs.

[13] provides an ontology and a set of basic formal definitions related to the comparison of RDF graphs. [14] and [15] describe two systems currently in development that allow for efficiently storing a potentially large number of variants of an RDF model by using a compact representation of the differences between them. These works concentrate on space-efficient storage and transmission of difference sets, but do not go into depth regarding how to use them to support higher-level tasks.

Finally, an extensive base of theoretical work is available from generic graph comparison research (see [16]), an area that is basically concerned with finding isomorphisms (or correspondences that approach isomorphisms according to some metric) between arbitrary graphs whose nodes and edges cannot be directly matched by name. This problem is analogous in many ways to the problem that interests us, but applies to a separate range of practical situations. In our case, we analyze the differences (and, of course, the similarities) between graphs whose nodes can be reliably matched in a computationally inexpensive way.

## 6 Summary and Future Work

Process model difference analysis helps to determine the differences between two variants of a process model, and offers flexible mechanisms to filter, analyze, and display those differences in specific ways, with the intent of supporting software process evolution. This type of analysis relies on the fact that the compared models contain a sizable common portion that can be used as a base for the comparison.

We have described two process management oriented scenarios where difference analysis can be used to support the tasks of many of the stakeholders involved in process improvement. The analysis of these scenarios allowed us to identify a number of concrete comparison operations that would arguably be useful while performing many of the discussed tasks.

Taking the scenarios and the particular comparison operation types into account, we discussed the main conceptual and technical challenges we think we have to overcome in order to implement a practical difference analysis system. We also presented a preliminary sketch of the software architecture for such a system.

Our aim is to completely implement a working difference analysis system, in order to validate its utility in practical scenarios. The main objectives for the validation are guaranteeing that our system allows us to specify a wide variety of useful comparisons with reasonable effort, and that the produced comparison results constitute useful support for the process improvement tasks at which they are targeted.

**Acknowledgments.** We would like to thank Sonnhild Namingha from Fraunhofer IESE for proofreading this paper. This work was supported in part by the German Federal Ministry of Education and Research (V-Bench Project, No.01| SE 11 A).